\documentclass[aps,prb,twocolumn,superscriptaddress,floatfix,showpacs]{revtex4}
\usepackage{graphicx}
\usepackage{epsfig}
\usepackage{amssymb}

\begin{document}

\title{Doniach diagram for ordered, disordered and underscreened Kondo lattices.}

\author{B. Coqblin}
\affiliation{Laboratoire de Physique des Solides, Universit\'{e}
Paris-Sud, UMR-8502 CNRS, 91405 Orsay, France}
\author{J. R. Iglesias}
\affiliation{Instituto de F\'{\i}sica, Universidade Federal do Rio
Grande do Sul, 91501-970 Porto Alegre, Brazil}
\author{N. B. Perkins}
\affiliation{Institute fur Theoretische Physik, TU Braunschweig,
38106 Braunschweig, Germany}
\affiliation{Bogoliubov Laboratory of Theoretical Physics, JINR,
141980 Dubna, Russia}
\author{S. G. Magalhaes}
\affiliation{Laboratorio de Mecanica Estatistica e Teoria da
Materia Condensada, Universidade Federal de Santa Maria,
97105-900 Santa Maria, Brazil}
\author{F. M. Zimmer}
\affiliation{Laboratorio de Mecanica Estatistica e Teoria da
Materia Condensada, Universidade Federal de Santa Maria,
97105-900 Santa Maria, Brazil}


\begin{abstract}
The Doniach's diagram has been originally proposed to describe the
competition between the local Kondo effect and the intersite RKKY
interactions in cerium compounds. Here we discuss the extension of
this diagram to different variations of Kondo lattice model.  We
consider a) ordered cerium compounds where the competition between
magnetic order and Kondo effect plays an important role, as in
$CeRh_2Si_2$, b) disordered cerium systems with competing spin glass
phase, magnetic ordered phases and a Kondo phase, as in the heavy
fermion cerium alloy $CeCu_xNi_{1-x}$ and, c) uranium compounds
where a coexistence between Kondo effect and ferromagnetic order has
been observed, as in UTe. We show that all these cases can be
described by a generalized Doniach phase diagram.
\end{abstract}



\maketitle

\section{Introduction}

The magnetism in strongly correlated $f$-electron systems is a
subject of great interest from both experimental and theoretical
points of view, since the famous explanation of the logarithmic
decrease of the electrical resistivity in magnetic dilute alloys
-- like AuFe -- by J. Kondo in 1964~\cite{Kondo}. This phenomenon
has been called Kondo effect, and in the following forty years it
has been experimentally observed in many heavy fermion compounds
with cerium, ytterbium, uranium and other rare earth or actinide
elements. An exact solution of the single impurity Kondo effect
has been theoretically determined by the renormalization group
and by Bethe ansatz:  in essence, the Kondo effect occurs because
the localized magnetic moment $S_f=1/2$ of the magnetic impurity
is completely screened by the conduction electron spin-density
``cloud'' at $T=0$. For a review see Ref~.\cite{Hewson,Coqblin1}.
However, the very rich phase diagram of dense heavy fermion
compounds can not be explained in the framework of the one
impurity model. It has been proven that the Kondo lattice (KL)
model is the appropriate tool for describing the properties of
many cerium and uranium compounds.

In heavy fermion materials there are two kinds of
electrons: conduction electrons from outer atomic orbitals, and
strongly correlated electrons from inner $f$-orbitals, the later
ones being generally localized. As a consequence, the $f$-electrons
may be considered as a lattice of localized spins and the KL model
describes the coupling of these localized spins with the electrons
in the conduction band. Historically, the KL model has been proposed
to account for properties of cerium compounds\cite{Iglesias,Coqblin3}. In most of
cerium-based compounds, cerium ions are in the $4f^1$ configuration.
In this case, there exists one $f$-electron per site, with spin $S=1/2$ spin,
which couples antiferromagnetically to the
conduction electron spin density, via an on-site exchange
interaction $J_K$. Besides, the local coupling between $f-$spins and
conduction electrons may give rise to magnetic order through the
RKKY interaction. The competition between Kondo effect, which is
characterized by an energy scale $T_K\sim exp(-1/J_K)$, and magnetic
order, characterized by the strength  of the RKKY interaction
$T_{RKKY}\sim J_K^2$, leads to a very rich phase diagram with
possible quantum phase transitions. In cerium compounds the
transitions typically occur between non-magnetic (Kondo) and
antiferromagnetically ordered metallic states. For small values of
$J_K$ coupling, the RKKY interaction is dominant and the system
orders magnetically. At intermediate values of $J_K$, energy scales
$T_K$ and $T_{RKKY}$ are of comparable strength -- magnetic order
still occurs but with partially screened localized moments. With
further increase of the $J_K$ coupling, the magnetic order is further
suppressed and only short range magnetic correlations can survive
\cite{Iglesias,Coqblin3} .

The presence of disorder in alloys can strongly affect the
competition between the RKKY interaction and  the Kondo effect.
The role of disorder  can be also studied in the framework of the
KL model if one considers the interaction between localized spins
as random variables with zero  mean. This extension is called
disordered Kondo lattice (DKL) model. In the DKL model disorder
produces a broad distribution of Kondo temperatures and can be
responsible for the deviation from the Fermi liquid behavior
found in some heavy fermion compounds. Indeed, various effects of
``Kondo disorder'' has been reported in  both cerium based --
CeNi$_{1-x}$Cu$_{x}$ \cite{Gomez-Sal1,Gomez-Sal2,Marcano},
Ce$_{2}$Au$_{1-x}$Co$_{x}$ Si$_{3}$ \cite{Majundar}--, or uranium
based -- $UCu_{5-x}Pd_{x}$ \cite{Volmer},
$U_{1-x}La_{x}Pd_{2}Al_{3}$ \cite{Zapf} -- compounds. Moreover,
most of these compounds exhibit coexistence of spin glass and
Kondo effect.

Finally, let us specially discuss the  behavior of uranium compounds, which
is generally quite different from cerium compounds, particularly due
to the existence of several coexistence phenomena between magnetism,
Kondo effect and
superconductivity~\cite{Schoenes,Schoenes3,Bukowski,Tran,Tran2,aoki,flouquet}.
The magnetism in uranium ions comes from $5f$ electrons, which in most
uranium compounds are in a crossover region between localized
and itinerant behavior. The strength of the localization depends
strongly on a subtle balance between electronic correlations,
crystal field and spin-orbit coupling. It is often rather difficult
to  decide, on the basis of the experimental data, between Kondo
behavior of well localized $5f^2$ configuration and mixed-valence
behavior. If the majority of uranium ions are in $5f^{2}$
configuration, in which two $f$-electrons are bound into spin $S=1$,
these uranium compounds can be studied in the framework
of the Underscreened Kondo lattice (UKL) model. This model considers
a periodic lattice of magnetic atoms with localized $S=1$ spins
interacting with conduction electrons via an intra-site
antiferromagnetic Kondo coupling and among them via an inter-site
ferromagnetic interaction.

The overall physics of heavy fermion compounds can be well
described by a generalized Doniach phase diagram. The original Doniach
phase diagram has been proposed to describe the experimentally
obtained phase diagram of some cerium compounds\cite{Coqblin3}.
In this phase diagram a quantum phase transition between a
magnetically ordered one phase and a non-magnetic Kondo phase is observed, i.e.
Kondo screening and RKKY coupling are in competition. The Doniach
diagram was later extended\cite{Iglesias,Coqblin3} to include the short range
magnetic correlations that survive inside the Kondo phase.

The competition between spin glass, magnetism, either ferro- or
antiferromagnetism, and Kondo effect can be also analyzed in the framework
of a Doniach phase diagram for the disordered Kondo lattice model.
In the case of antiferromagnetism,
the obtained phase diagram shows the sequence of spin glass,
antiferromagnetic order and  Kondo state for  increasing Kondo
coupling $J_K$\cite{Maga2,Magalhaes}. The experimental phase diagram
of Ce$_{2}$Au$_{1-x}$Co$_{x}$ Si$_{3}$ \cite{Majundar} can be
addressed if the $J_K$ coupling is associated with the concentration
of cobalt. In the case of ferromagnetism, the phase diagram displays several
phase transitions among a ferromagnetically ordered region, a spin
glass one, a mixed phase and a pure Kondo state\cite{Magalhaes2}.
This phase diagram could be used to partially address the experimental
data for CeNi$_{1-x}$Cu$_{x}$ \cite{Gomez-Sal1,Gomez-Sal2}.

An analog of the Doniach phase diagram can be derived for the UKL
model, and it appears to be significantly different from the
original one for the Kondo lattice model with localized spins
$S=1/2$. When decreasing the temperature, the UKL model exhibits two
continuous transitions (more precisely, sharp crossovers): the first
one, at $T=T_K$, to a non-magnetic Kondo state, and the second one,
at $T=T_C$, to a ferromagnetic state that coexists with the Kondo
state. At $T=0$, for a strong Kondo coupling, $J_K$, ferromagnetism
and Kondo effect coexist. The obtained phase diagram has regions of
Kondo-ferromagnetic coexistence, non-magnetic Kondo behavior and
pure ferromagnetism. This phase diagram can be considered as a new
ferromagnetic Doniach diagram for the UKL model.

In this review we analyze how the interplay  between Kondo effect,
magnetic order, antiferromagnetic or ferromagnetic, and spin glass
behavior modify the Doniach phase diagram for the Kondo lattice. We argue
that the Doniach phase diagram is a useful theoretical tool for
studying such competition, and, at the same time, provides  a
convenient playground for experimentalists to study the stability of
magnetic order in concentrated rare earth and actinide based
compounds.

\section{S=1/2 Kondo lattice model: Cerium compounds.}

In a large number of cerium compounds, Ce ions are in the localized $4f^1$
configuration corresponding to spin $S=1/2$. The localized spin
couples antiferromagnetically, via an on-site exchange interaction,
$J_K$, to the conduction electron spin density.  At very low
temperatures the localized spin $S=1/2$ is completely screened by
the conduction electrons, leading to the formation of coherent Kondo
spin-singlet state. Besides, the local coupling between $f-$spins
and conduction electrons may give rise to a magnetic order through
the RKKY interaction.

Most cerium compounds exhibit antiferromagnetic correlations between
Ce ions, and many of them order antiferromagnetically, with rather
low N\'eel temperatures, about 10 K. At low temperatures there is a
strong competition between the Kondo effect and antiferromagnetic
order. This competition has been firstly described by the well-known
Doniach diagram. Accordingly, cerium compounds are separated into
those which do not order magnetically and have a very large
electronic specific heat constant $\gamma$ (like CeAl$_3$), and
those which order magnetically and present a relatively smaller
heavy fermion behavior at low temperatures (like CeAl$_2$). Detailed
reviews and references can be found in refs.
\cite{Coqblin1,Iglesias}. On the other hand, there are very few
compounds which exhibit both a Kondo character and a ferromagnetic
order (generally with a low Curie temperature). For example, a
recently studied compound is CeRuPO which order ferromagnetically at
$T_C =15 K$ and presents a relatively weak Kondo effect
\cite{Krellner}.

The Hamiltonian of the KL model can  be written as:
\begin{eqnarray}
\begin{array}{l}
H = \sum_{k\sigma}\epsilon_{k} n^{c}_{k\sigma}+
\sum_{i\sigma}\epsilon_0 n^{f}_{i\sigma} +J_K \sum_{i}{\bf
S}_i{\bf{\sigma}}_i \\ + \frac{1}{2}J_H \sum_{ij}{\bf S}_i{\bf
S}_j\\\label{Ham}
\end{array}
\end{eqnarray}
 \noindent where the first term represents the
conduction band with dispersion $\epsilon_{k}$, width $2D$ and
constant density of states $1/2D$; $\epsilon_{0}$ is a Lagrange
multiplier which is fixed by a constraint for the total number of
f-electrons per site, $n_{f}=1$, and can be interpreted as a
fictitious chemical potential for $f$-fermions. The third term is
the on-site antiferromagnetic Kondo coupling, $J_{K}>0$, between
localized $\mathbf{S_i}=1/2$ and conduction electron's
$\sigma_i=1/2$ spins. Finally, the last term is the
antiferromagnetic inter-site interaction, $J_{H}>0$, between
localized $f$-magnetic moments.
\begin{figure}[h]
\centerline{\includegraphics[width=8.cm, clip=true]{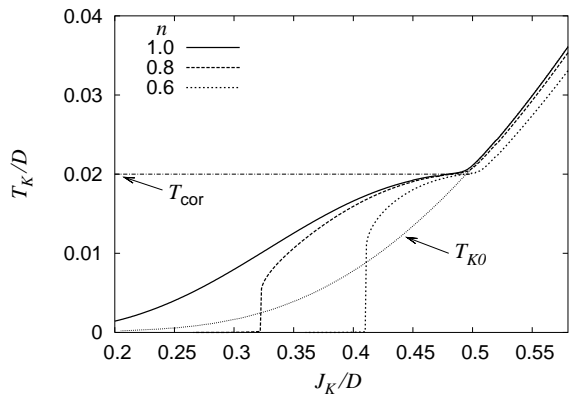}}
\caption{The non-magnetic region of the Doniach diagram for Ce compounds: Plot of the
Kondo temperature $T_{K}$ as a function of $J_{K}$ for $J_{H}/D=0.04$
and representative values of the conduction band filling, $n$. It is also showed
the correlation temperature $T_{\rm cor}$, and the single-impurity Kondo temperature $T_{\rm K0}$.
Reproduced from ref.~\cite{Coqblin3}} \label{KLI}
\end{figure}

This Hamiltonian has been studied within a generalized mean field
scheme where the relevant order parameter related to the Kondo
effect is $\lambda_{\sigma}=\langle
c_{i\sigma}^{+}f_{i\sigma}\rangle$. A finite value of $\lambda$
corresponds to a Kondo state (for a full discussion of the limits
and validity od this approximation see
references~\cite{Iglesias,Coqblin3,Perkins}). Within this mean field
treatment a non-exponential Kondo temperature can be obtained, and
also a correlation critical temperature for the short-range magnetic
correlations near the quantum critical point. In Fig.~\ref{KLI} we
show the results of the Kondo temperature as a function of $J_K$ in
the non-magnetic region. It is evident that for intermediate values
of $J_K$ the Kondo temperature clearly deviates from the
one-impurity exponential behavior, and also a short range magnetic
correlation between neighboring spins is present, which is
characterized by a correlation temperature $T_{cor}$. For high
values of the interaction $J_K$ the Kondo temperature becomes
independent of the electron concentration and almost equal to the
one-impurity value. These results account for the behavior of $T_K$
in some cerium compounds where a deviation from the exponential law
has been observed~\cite{Coqblin3,Iglesias}. We remark that in the KL
model there is no coexistence between the Kondo region and the
magnetic ordered phase. If one includes possible magnetic solution,
a narrow region of coexistence is obtained, but either the Kondo or
the magnetic solution is the minimum free energy one, it was
discussed in ref.~\cite{Gusmao}). Because of this reason the
magnetic region, corresponding to low values of $J_K$, is not
depicted in Fig.~\ref{KLI}.

\section{The spin glass  and magnetism in Kondo lattice.}

The second theoretical study we present here concerns the phase
diagram of the disordered KL the model. The DKL model is an
extension of KL model with a random intersite interaction between
the localized spins. The Hamiltonian describing this situation is given by:
\begin{eqnarray}
\begin{array}{l}
H=\sum_{k\sigma}\epsilon_{k} n^{c}_{k\sigma}+
\sum_{i\sigma}\epsilon_0 n^{f}_{i\sigma} +H_{SG}\\ +J_{K}\sum_{i}
({S}^{+}_{i}{s}^{-}_{i} +h.c.)\\ \label{e2}
\end{array}
\end{eqnarray}
where the term $H_{SG}=\sum_{i,j}J_{ij}{S}^{z}_{i}{S}^{z}_{j}$ is
given by a quantum Ising interaction between the $z$-components of
the localized spins, and it describes a spin glass character. There
are several ways  how one can introduce randomness into exchange
integrals $J_{ij}$. First, the  exchange integrals can be considered
as random variables with a Gaussian distribution with zero average,
like in the Sherrington-Kirkpatrick model ~\cite{SK}. In this case
the obtained phase diagram shows first a spin glass phase and then a
Kondo phase with increasing $J_K$ ~\cite{Alba}. Important
improvements can be achieved when a non zero average value of the
Gaussian distribution is considered. Then  a more complex phase
diagram with a ferromagnetic ~\cite{Maga1} or an antiferromagnetic
~\cite{Maga2} phase occurs at low temperatures for smaller $J_K$
values. This approach allows  also to derive a quantum critical
point for the magnetic phase ~\cite{Magalhaes,Alba2}.

Here we discuss some recent theoretical developments which allows to
better understand the spin glass and Kondo state coexistence which
have been experimentally observed in several cerium
~\cite{Gomez-Sal1,Gomez-Sal2} and uranium ~\cite{Volmer,Zapf}
alloys. This approach is a generalization of the Mattis model
~\cite{Mattis}, and it represents an interpolation between a
ferromagnet and a highly disordered spin glass (Detailed
calculations can be found in Ref. ~\cite{Magalhaes2}). The coupling
between spins on different sites, $i$ and $j$, is given by:
\begin{equation}
J_{ij}=\frac{1}{N}\sum_{\mu}J\xi_i^{\mu}\xi_j^{\mu} \label{Jij},
\end{equation}
where the $\xi_i^{\mu}=\pm 1$ ($\mu=1,2,...,p$; $i=1,2,...,N$) are
independent randomly distributed variables. For the
classical Ising model, if $\mu=1$, the original Mattis model~
\cite{Mattis} is recovered. However, if p=N  and the $N^{2}$
random variables $\xi_i^{\mu}$ have zero mean and unitary variance,
, $J_{ij}$ tends to a Gaussian variable in the limit of N large, as in the
Sherrington-Kirkpatrick model \cite{SK}.
Therefore, one can consider this model as an interpolation between
ferromagnetism and highly disordered spin glass.

The critical parameter of the model is the ratio $a=p/N$, which
describes the degree of the frustration induced by disorder. For
large values $a>a_c$, where $a_c$ is a  particular value, the
frustration is dominant leading to a spin glass behavior,
characterized by a certain temperature $T_f$. At small value of
$a<a_c$, this parameter gives an estimation of the relative
importance of the ferromagnetic and spin glass phases for small
$J_K$ values, because  for large $J_K$ values, the Kondo phase is
always present. In fact, for low values of both $a$ and $J_K$ a complex
behavior can appear. Therefore, in the presence of disorder, the
parameter $a$ together with $J_K/J$ constitute the parameter space
of the extended Doniach phase diagram of the DKL model.

\begin{figure}[h]
\centerline{\includegraphics[angle=-90,width=8.cm,
clip=true]{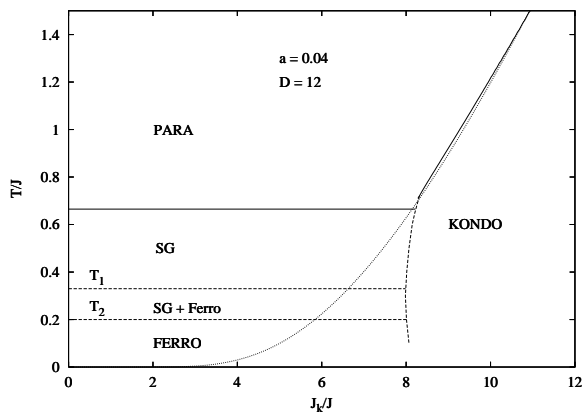}} \caption{The Kondo-spin
glass-ferromagnetic diagram versus $J_K/J$ for D=12 and a=0.04.}
\label{Spinglass}
\end{figure}

In Fig.~\ref{Spinglass} we present the phase diagram for $a=0.04$.
For a relatively small $J_K/J$ ratio and when decreasing the temperature,
there is a a spin glass phase, then a coexistence region between
ferromagnetism and spin glass and, finally, a ferromagnetic phase.

Fig.~\ref{Spinglass} accounts very well for the experimental phase
diagram of $CeCu_{1-x}Ni_x$ disordered alloys, where Kondo, spin
glass and magnetically ordered phases are observed. The experimental
phase diagram of $CeCu_{1-x}Ni_x$ is in fact very complicate: for
example,a very careful experimental study of $CeCu_{0.6}Ni_{0.4}$
yields a percolative transition with decreasing temperature from a
cluster-glass state with ferromagnetic correlations to a
ferromagnetic clustered state and recent theoretical simulations can
reproduce satisfactorily the experimental situation~\cite{Marcano}.
But the results presented in Fig.~\ref{Spinglass}, and in a more
detailed way in Ref.\cite{Magalhaes2}, provide a fairly good description
of the phase diagram of $CeCu_{1-x}Ni_x$ alloys.

\section{The underscreened Kondo lattice applied to uranium compounds}

As we already said in the introduction, uranium compounds show very
rich behaviors, quite different from cerium compounds, with the
presence of numerous coexistence phenomena. Here we discuss mainly
the coexistence of magnetic order with Kondo effect, which can be
well described in the framework of a extended ferromagnetic Doniach diagram.
Let us briefly describe the experimental situation. The first
experimental evidence of the coexistence between Kondo behavior and
ferromagnetic order in the dense Kondo compound $UTe$ has been
obtained long time ago~\cite{Schoenes}. More recently, this
coexistence has been observed in $UCu_{0.9}Sb_{2}$ ~\cite{Bukowski}
and $UCo_{0.5}Sb_{2}$ ~\cite{Tran,Tran2}. All these systems undergo
a ferromagnetic ordering at the relatively high Curie temperatures
of $T_{C}$ = 102K ($UTe$), $T_{C}$ = 113K ($UCu_{0.9}Sb_{2}$) and
$T_{C}$ = 64.5K ($UCo_{0.5}Sb_{2}$). Above the ordering
temperatures, i.e. in the expected paramagnetic region, these
materials exhibit a Kondo-like logarithmic decrease of the
electrical resistivity, indicating a Kondo behavior. This
logarithmic variation extends down to the ferromagnetic Curie
temperature, $T_C$, suggesting that the Kondo behavior survives
inside the ferromagnetic phase, implying that the ferromagnetic
order and the Kondo behavior do coexist. This coexistence, together
with the large Curie temperatures, are clearly novel features that
cannot be explained by the standard KL model.

This coexistence can be qualitatively  understood by studying the
underscreened Kondo lattice model. Detailed results can be found in
refs.~\cite{Perkins,Perkins02}. The UKL  model describes the situation when the
localized spins are larger than  $1/2$,  and therefore, they cannot
be completely screened by the conduction electrons at $T=0$. The UKL
model  can be applied to uranium compounds if the majority of
uranium ions are in $5f^{2}$ configuration, in which two
$f$-electrons are bound into spin $S=1$. We believe this is an
appropriate description of electronic states for those compounds
which have magnetic moments close to the free ion
values~\cite{Schoenes,Bukowski,Tran}. Here we will describe a mean
field solution of the UKL model, which  describes well the physics
of the above mentioned uranium compounds.

The Hamiltonian of the UKL model can  be  formally written as
Eq.(1), the one for the normal KL model. The main difference between the KL Hamiltonian (Eq.(1)) and the UKL model
is that the localized spins are now $\mathbf{S_i}=1$. Correspondingly,
the constraint for the total number of f-electrons per site is
modified, and it is given by $n_{f}=\sum_{\sigma}(n_{\sigma}^{f1}+n_{\sigma}^{f2})=2$, where $f1$
and $f2$ are two degenerate $f$-orbital states. Also, the last term of the Hamiltonian now represents a
ferromagnetic inter-site interaction between localized $f$-spins, then for the UKL Hamiltonian $J_{H}<0$.

We restrict our considerations to a mean-field (MF) treatment of
the Hamiltonian and introduce four MF parameters
$\lambda_{\sigma}=\langle\sum_{\alpha}
c_{i\sigma}^{+}f_{i\sigma}^{\alpha}\rangle$, $M =\frac{1}{2}\langle
{n_{i\uparrow}^f-n^f_{i\downarrow}}\rangle$ and
$m=\frac{1}{2}\langle {n^c_{i\uparrow}-n^c_{i\downarrow}}\rangle$,
 where $\langle...\rangle$ denotes the thermal average.

\begin{figure}[h]
\centerline{\includegraphics[width=8.cm, clip=true]{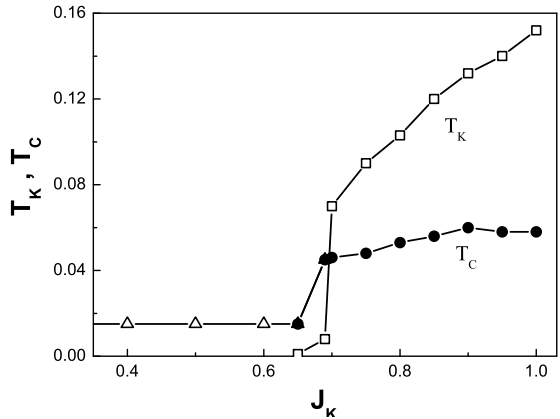}}
\caption{The ferromagnetic Doniach diagram: Plot of the Curie
temperature $T_C$ and the Kondo temperature $T_K$ versus $J_K$ for
$J_H= -0.01$ and $n_c= 0.8$. Reproduced from Ref.~\cite{Perkins}}
\label{Doniach}
\end{figure}

Non-zero values of  $M$ and $m$ describe a magnetic phase with
non-zero total magnetization, while non-zero values of
$\lambda_\sigma$ describe the Kondo effect and the formation of the
heavy-fermion state, as in the case of the standard KL. For large
values of $J_K$, one obtains a Kondo-ferromagnetic state below the
Curie temperature $T_c$, then a Kondo state between $T_c$ and the
Kondo temperature $T_K$ and finally a paramagnetic state above
$T_K$. Then, the critical temperatures, $T_c$ and $T_K$, have been
derived as a function of $J_K$. We would like to stress that $T_K$
increases abruptly above a critical value $J_K^c$, while $T_c$ is
always non zero but increases slowly above $J_K^c$ and then remains
at a value smaller than $T_K$. This ferromagnetic-Kondo diagram is
shown in Fig.~\ref{Doniach} and is completely different from the
Doniach diagram derived for cerium compounds. Thus, the UKL model
provides the first explanation of the behavior of ferromagnetic
Kondo uranium compounds, describing the coexistence of Kondo
behavior and ferromagnetic order.

Besides its applicability to the physics of ferromagnetic uranium
compounds, the UKL model is an interesting problem on its own, but
yet it has attracted a little attention. Possible extensions of this
model would be very interesting, in order to describe the
coexistence between the ferromagnetic order and Non-Fermi-Liquid
behavior, as observed in $URu_{2-x}Re_{x}Si_{2}$
compounds\cite{Bauer} or the decrease of  $T_c$
versus $J_K$ down to a eventual quantum critical point.

\section{Conclusions}

We have analyzed how the interplay  between Kondo effect,
magnetic order, (antiferromagnetic or ferromagnetic), and spin glass
behavior modify the Doniach phase diagram for the Kondo lattice.
We have discussed in detail the cases of a) cerium compounds
where the Kondo effect coexists with short range magnetic correlations
but competes with long range magnetic order,
b) disordered cerium alloys of the type $CeCu_xNi_{1-x}$, where the
obtained Doniach diagram exhibits magnetic ordered, spin-glass and Kondo
phases and c) uranium compounds, like $UTe$, which
are well described by the UKL model, that can account for the observed coexistence
between ferromagnetic order and Kondo behavior observed in those uranium compounds.


\end{document}